\documentclass[notitlepage,superscriptaddress,twocolumn,preprint]{revtex4-1}
\usepackage{graphicx}
\usepackage{amsmath,amssymb}
\usepackage{tabularx}
\usepackage[breaklinks,colorlinks,linkcolor=blue,citecolor=blue,urlcolor=blue]{hyperref}
\usepackage{MnSymbol}
\usepackage{overpic}
\usepackage{subfig}

\begin{document}

\title{Antiferromagnetic phase transition in the temperature-dependent NIR-VUV dielectric function of hexagonal YMnO$_3$}

\author{Steffen Richter}
\affiliation{Universit\"at Leipzig, Institut f\"ur Experimentelle Physik II, Linn\'estr. 5, 04103 Leipzig, Germany}

\author{Stefan G. Ebbinghaus}
\affiliation{Martin-Luther-Universit\"at Halle-Wittenberg, Institut f\"ur Chemie, Kurt-Mothes-Str. 2, 06120 Halle, Germany}

\author{Marius Grundmann}
\affiliation{Universit\"at Leipzig, Institut f\"ur Experimentelle Physik II, Linn\'estr. 5, 04103 Leipzig, Germany}

\author{R\"udiger Schmidt-Grund}
\affiliation{Universit\"at Leipzig, Institut f\"ur Experimentelle Physik II, Linn\'estr. 5, 04103 Leipzig, Germany}

\date{Feb 2015}

\begin{abstract}
Hexagonal YMnO$_3$ is well known for the co-occurrence of ferroelectricity and antiferromagnetism at low temperatures. Using temperature-dependent spectroscopic ellipsometry at an $a$-plane oriented single crystal, we show how the dielectric function is affected by the magnetic order transition at the N\'eel temperature. 
We focus especially on the pronounced charge transfer transitions around ($1.6-1.7$)\,eV which are strongly connected to Mn 3$d$ electrons. 
If described with a Bose-Einstein model, the temperature dependency of their energy and broadening is characterized by effective phonon energies not larger than 8\,meV. We argue that this is a hint for the occurrence of a soft phonon mode related to the antiferromagnetic phase transition. 
This is observed in both tensor components of the dielectric function, parallel and perpendicular to the crystallographic $c$-axis. 
Furthermore, a suitable parametrization for the uniaxial dielectric function is presented for the NIR-VUV spectral range. The broad transitions at energies higher than a critical point-like bandgap do not show a clear temperature dependence. We also observe some weak discrete absorption features around the strong charge transfer transitions with energies matching well to low-temperature photoluminescence signals. 

\end{abstract}

\maketitle

Rare earth manganites \textit{R}MnO$_3$ are interesting due to the occurrence of multiferroic properties. 
The co-existence of ferroelectricity and magnetic ordering in hexagonal (h) YMnO$_3$ has been known since the 1960s \cite{smolenskii}. 
h-YMnO$_3$ is ferroelectric up to a ferroelectric Curie temperature in the order of 920K \cite{abrahams01}. In addition it reveals antiferromagnetic ordering below the N\'eel temperature $T_N$. Values for $T_N$ between 60 and 90K have been reported, most of them slightly above 70K \cite{tomuta,katsufuji,froehlich, huang}. 
YMnO$_3$ occurs in orthorombic and hexagonal phases which both are ferroelectric but due to different origins \cite{cheong,zhou}. In ferroelectric h-YMnO$_3$ (space group $P6_3cm$), the electric dipole along the $c$-axis originates from a buckling of the MnO$_5$ bipyramid layers and a related displacement of the Y$^{3+}$ ions \cite{vonaken}. 
The antiferromagnetism at low temperatures is caused by frustration of the $S=2$ spins of the Mn$^{3+}$ ions which are arranged in a triangular lattice within the MnO$_5$ layers ($ab$-plane) \cite{tomuta,pailhes}. However, an additional ferromagnetic component along the $c$-axis can exist \cite{singh}. 
Magneto-electric coupling is a consequence of the configurational interplay of the manganese ions and oxygen ions. 

Besides studies of the antiferromagnetic phase transition by temperature-dependent magnetic susceptibility or specific heat \cite{tomuta, katsufuji, zhou} 
there has been a number of investigations utilizing (low-frequency) dielectric constant determination \cite{tomuta, katsufuji,huang}, neutron scattering \cite{pailhes, lee} and even optical second harmonic generation \cite{froehlich, pisarev}, all showing a distinct anomaly at the antiferromagnetic phase transition. This indicates a strong coupling between the magnetic order and the electronic system. 
For similar hexagonal \textit{R}MnO$_3$ systems there exist also IR and THz investigations showing that certain phonon modes are affected by the antiferromagnetic phase transition and in the antiferromagnetic phase spin lattice waves (magnon excitations) occcur \cite{liu,standard}. 
All those effects indicate strong charge-spin interactions, i.e. connection between magnetic and electronic systems. 

Recently, we have determined the highly anisotropic, uniaxial dielectric function (DF) of h-YMnO$_3$ at room temperature in a wide spectral range from 0.5 to 9.15\,eV \cite{ymo1}.
Outstanding are the distinct peak-shaped transitions at ($1.6-1.7$)\,eV, which are connected to an inter-site charge transfer transition from hybridized O 2$p$ electron orbitals to Mn 3$d$ states \cite{choi08prb77}. Thus these electronic resonances are closely connected to the orbitals involved in the magnetic ordering.

So far, there is only little literature about the temperature- and magnetic field dependency of electronic resonances in the UV-VIS spectral range. For YMnO$_3$ there is one photoluminescence (PL) study on a thin film showing that the emission around 1.7\,eV increases drastically when entering the antiferromagnetic phase \cite{nakayama}. 
For other hexagonal manganites temperature- and magnetic field behavior of such charge transfer transitions have been studied by transmission measurements proving indeed a strong connection to the magnetic order \cite{choi08prb78}. 
However, nearly all present investigations lack of considering the anisotropy since usually only polarization perpendicular to the crystallographic $c$-axis is probed.

\begin{figure*}
	\centering
	\subfloat{
		\begin{overpic}
			[height=.385\textwidth]	{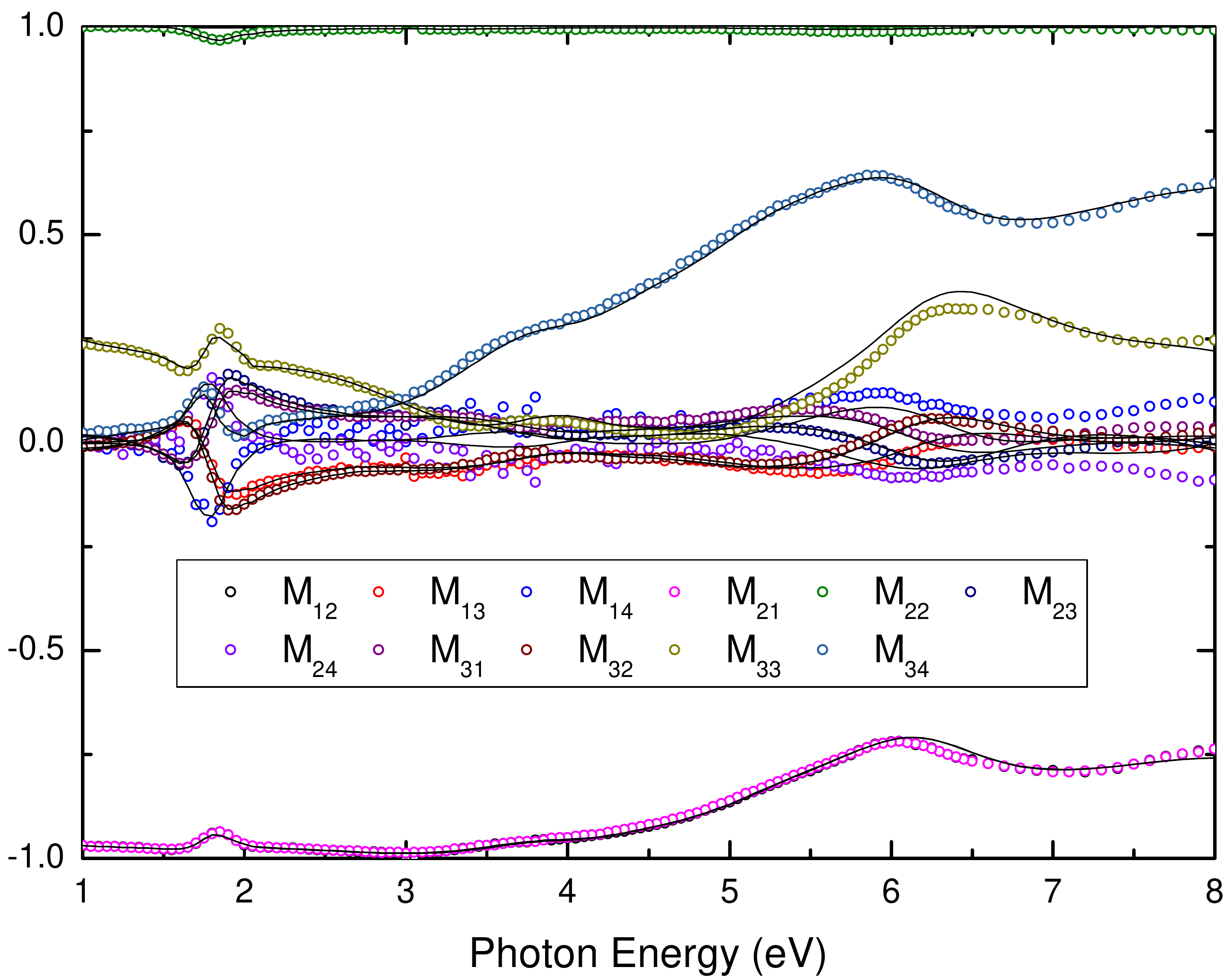}
			\put(3,1){a)}
		\end{overpic}
    }
	\subfloat{
		\begin{overpic}
			[height=.385\textwidth]	{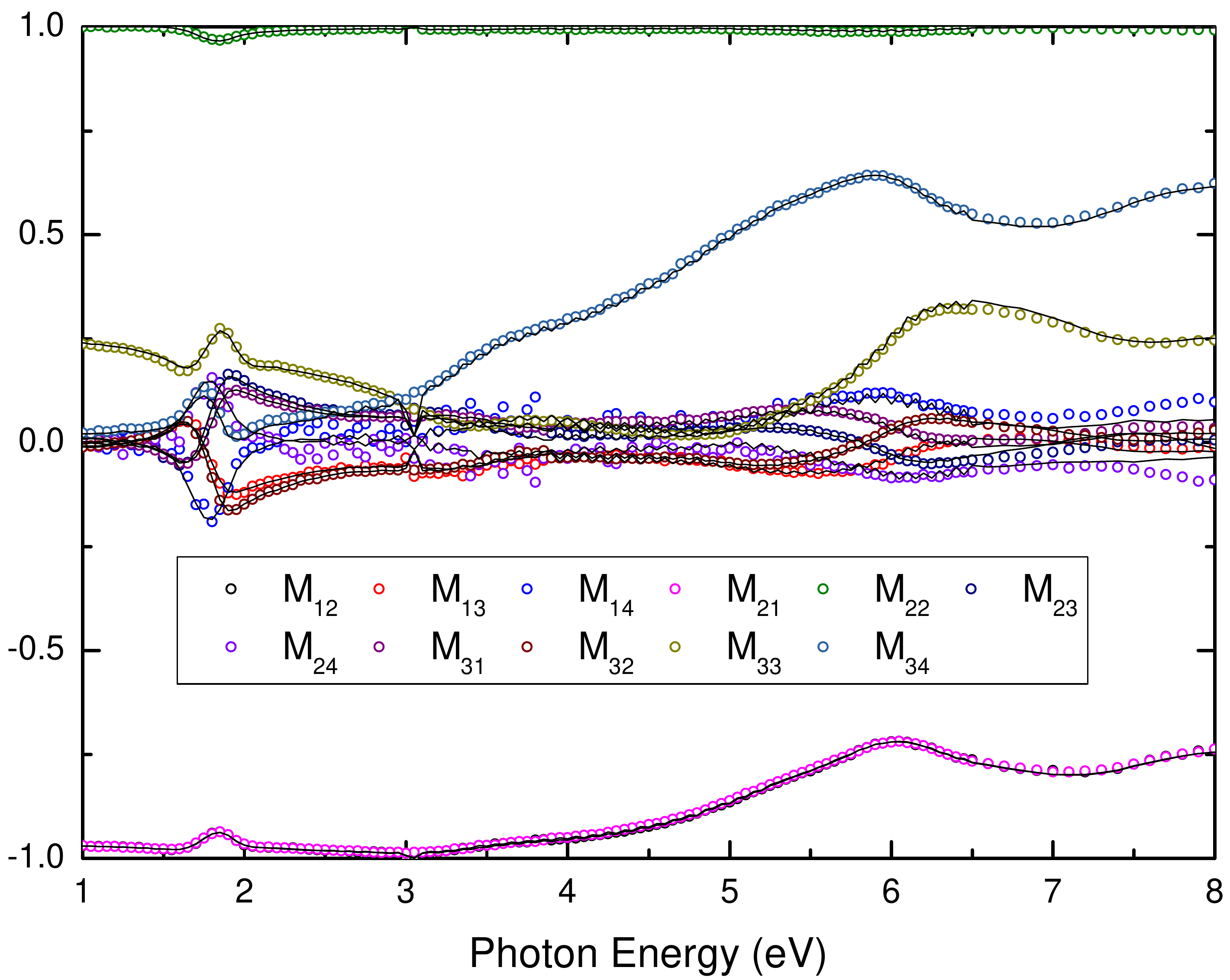}
			\put(3,1){b)}
		\end{overpic}
	}
\caption{Experimental and model-generated M\"uller matrix elements for $T$=10K: a) Using parametric oscillator model-DF, b) Using a wavelength-by-wavelength numeric model-DF. The incidence angle was 70$^\circ$ and the sample was orientated with an angle of 30$^\circ$ between the crystallographic $x$-axis and the plane of incidence.} 
	\label{fig:datenanpassung}
\end{figure*}

\begin{figure*}
	\centering
	\subfloat{
		\begin{overpic}
			[height=.375\textwidth]	{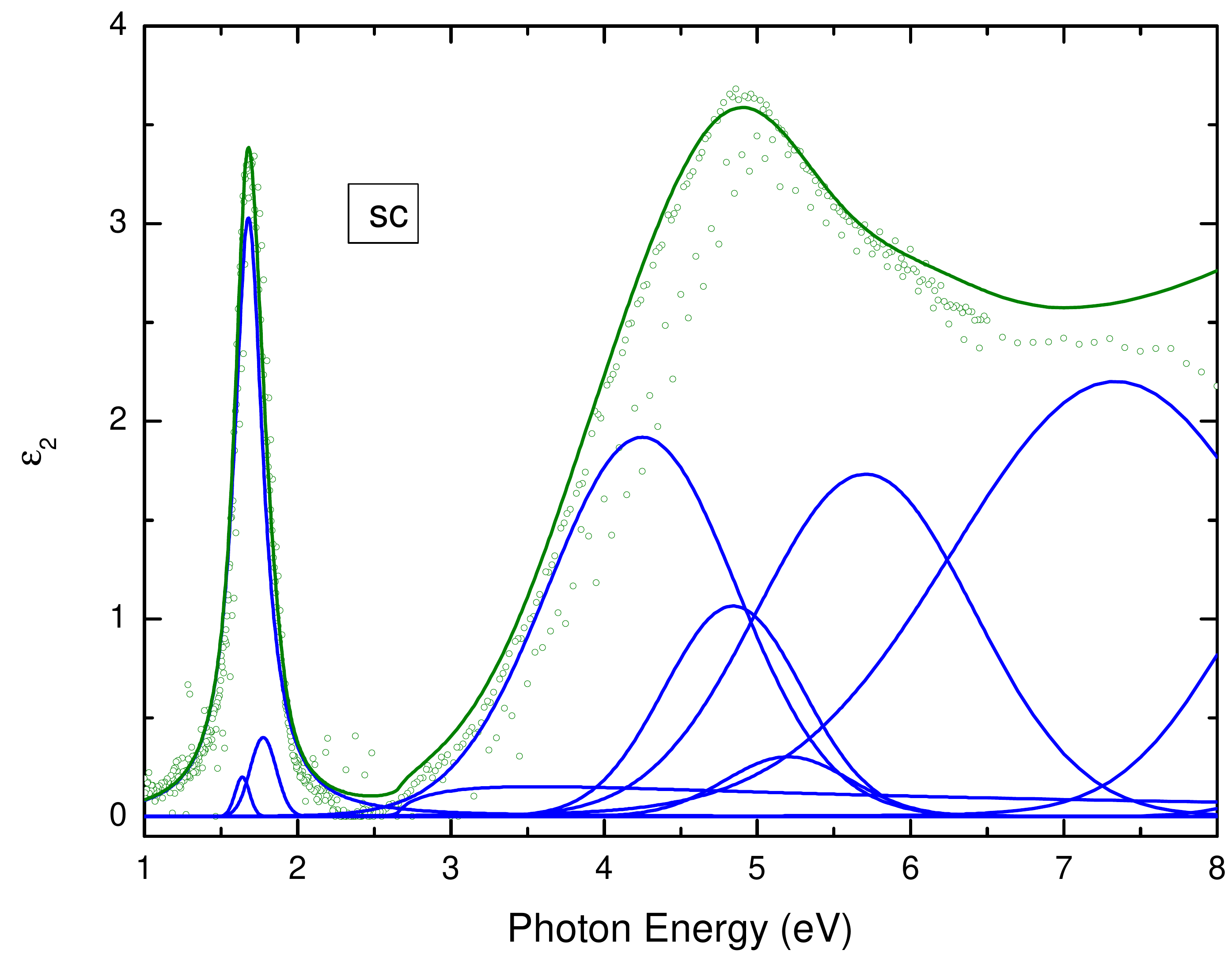}
			\put(3,1){a)}
		\end{overpic}
    }
	\subfloat{
		\begin{overpic}
			[height=.375\textwidth]	{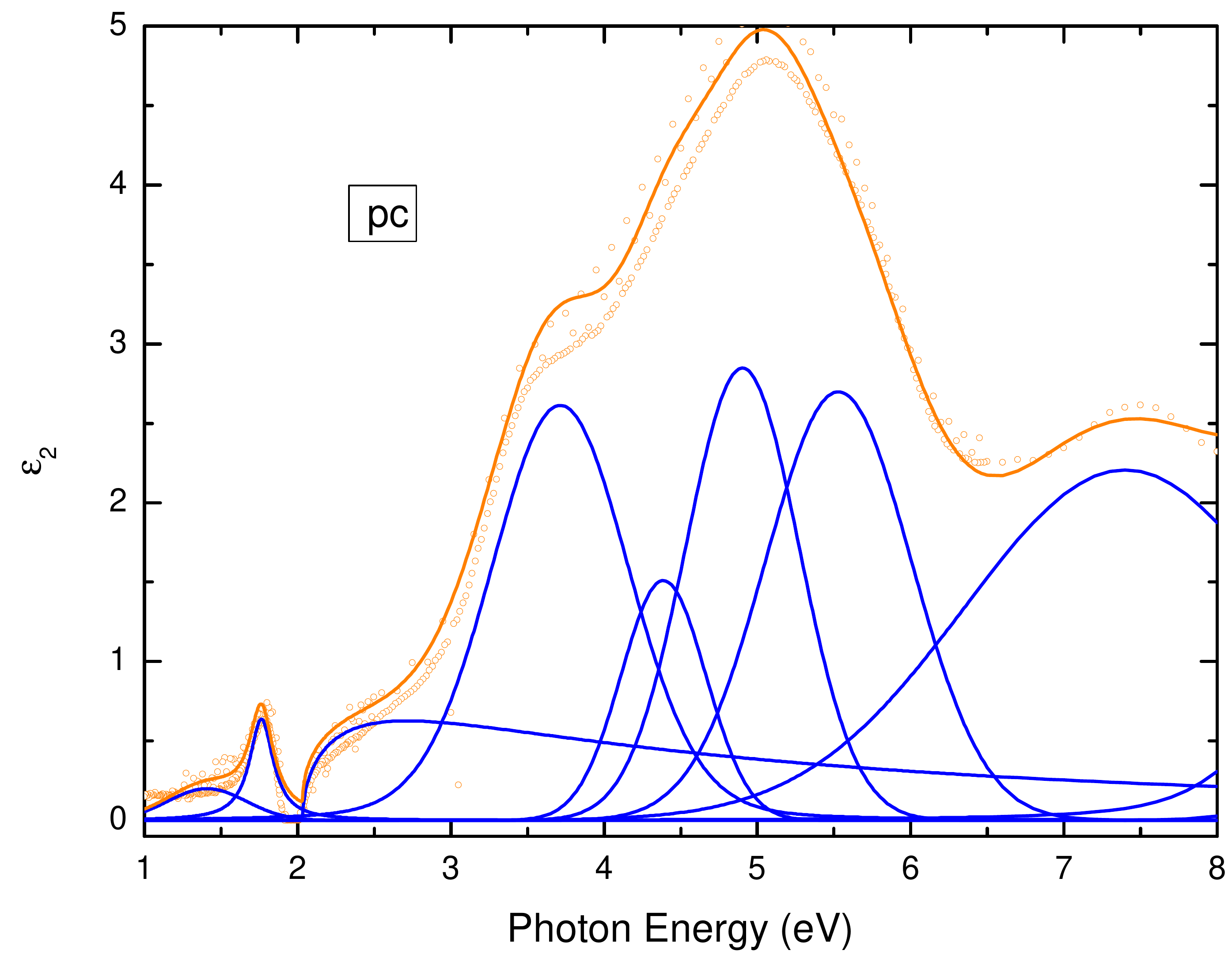}
			\put(3,1){b)}
		\end{overpic}
    }
	\caption{Imaginary part of the parametric model-DF for YMnO$_3$ at $T=$10K for a) sc and b) pc polarizations showing the individual contributions obtained by applying the parametric oscillator model-DF in blue lines (see fig. \ref{fig:datenanpassung}). Circles represent the wavelength-by-wavelength numeric model-DF.}
	\label{fig:dfcontributions}
\end{figure*}

\section*{Experimental details}

The single crystal sample is $a$-plane oriented with a slight miscut as known from previous investigations \cite{ymo1}. It was grown by the optical floating zone technique. 
Temperature-dependent spectroscopic ellipsometry  measurements were carried out from 10K to room temperature at an incidence angle of 70$^{\circ}$ using a cryostat  with CaF$_2$ windows. To account for polarization influences from the cryostat windows we applied a calibration procedure considering the window effects in terms of system Jones matrices (cf. \cite{tobias}). For each wavelength 12 of the 16 M\"uller matrix elements could be measured independently. 
Two different sample orientations with respect to the orientation of the crystallographic $c$-axis were measured to properly distinguish the polarizations parallel (pc) and perpendicular (sc) to the $c$-axis in the evaluation (cf. \cite{ymo1}). 
During the measurements the pressure in the cryostat was kept below $5\times 10^{-8}$\,mbar. Yet below 110K growing of a very thin ice layer was observed. To include this in the evaluation and reliably determine the DF of YMnO$_3$ the assumed sample model consists of uniaxial bulk material covered by an anisotropic surface layer. 
The latter was implemented as effective medium approximation according to Bruggeman \cite{bruggeman} with mixing of the optical constants of YMnO$_3$, ice and vacuum in varying amounts. Hence it accounts for roughness and possible nucleation of ice. Furthermore, only partial coverage was assumed to model also slight depolarization. The thickness of this surface layer is in the order of 6nm which is consistent with the sample topology investigated by atomic force microscopy as discussed in \cite{ymo1}. 

The obtained YMnO$_3$-DF is very sensitive to the specific surface treatment. 
In order to get a more robust evaluation and gain statistics, two different approaches were applied: 
A) A parametric model-DF based on oscillator functions was directly used to simulate the experimental data. 
B) A numeric model-DF was extracted wavelength by wavelength to simulate the experimental data. The numeric DF values were then approximated by the parametric model-DF. The different approaches can be seen in figures \ref{fig:datenanpassung} and \ref{fig:dfcontributions}. 
For further comparison of the model's robustness the spectral range was restricted to ($1-3$)\,eV and ($1-8$)\,eV, respectively, during modeling.

\section*{Results and discussion}

\begin{table*}
	\centering
		\caption{Dielectric function contributions with mean parameter values averaged over all temperatures.}
		\vspace{.5cm}
		\begin{tabular}{c c c c c}
		polarization & oscillator & $E$ & FWHM & amplitude \\
			& type & eV & eV &  \\
			\hline
			pc & Gaussian & 1.4 & 0.6 & 0.2 \\
			& Lorentzian & see below & see below & 0.6 \\
			& M0-CP \cite{adachi} & 2.0 & 0.01 & 5\,(eV)$^{\frac{3}{2}}$ \\
			& Voigt & $3.65$ & $1.0$ & $2.4$ \\
			& Gaussian & $4.35$ & $0.7$ & $1.8$ \\
			& Gaussian & $4.9$ & $0.9$ & $3.0$ \\
			& Gaussian & $5.55$ & $1.1$ & $2.7$ \\
			& Voigt & $7.4$ & $2.5$ & $2.3$ \\
			& $\varepsilon_\infty$=1.64 & & &  \vspace{.25cm} \\
			\hline 
			sc & Gaussian & 1.75 & 0.2 & 0.4 \\
			& Lorentzian &  see below & see below & 3.0 \\
			& Gaussian & 1.57 & 0.1 & 0.2 \\
			& M0-CP \cite{adachi} & 2.6 & 0.02 & 2\,(eV)$^{\frac{3}{2}}$ \\
			& Voigt & $4.3$ & $1.4$ & $1.9$ \\
			& Gaussian & $4.9$ & $0.8$ & $1.1$ \\
			& Gaussian & $5.4$ & $0.9$ & $0.4$ \\
			& Gaussian & $5.7$ & $1.7$ & $1.7$ \\
			& Voigt & $7.5$ & $2.5$ & $2.5$ \\
			& $\varepsilon_\infty$=1.55 & & &  \\
		\end{tabular}
	\label{tab:parametrization}
\end{table*}

\textit{Dielectric function parametrization}

Starting from the well-known numeric DF for both polarizations \cite{ymo1}, a simplified general oscillator model was developed as parametrized model-DF: For each polarization it consists of a M0-like critical point transition ('direct bandgap'), five Lorentzian, respectively Gaussian, functions above this gap and the distinct Lorentzian transitions around ($1.6-1.7$)\,eV with one (pc) or two (sc) weak Voigt-shaped resonances close to it (cf. fig. \ref{fig:dfcontributions} and tab. \ref{tab:parametrization}). 
As the pronounced charge transfer occurs from hybridized O $2p_{x/y}$ orbitals to Mn $3d_{z^2}$ orbitals, those weak transitions arise most likely due to similar transitions but outgoing from differently hybridized states of O $2p_{x/y}$ with Mn $3d_{xy/yz}$ orbital contributions. Such weaker transitions are observed in other hexagonal manganites as well (cf. \cite{choi08prb77}). 
In sc polarization such a weak resonance is also found at an energy slightly higher than that of the main resonance while in pc polarization a weak oscillator is only found at lower energies with a much larger broadening compared to the sc polarization. 
The occurrence of critical point-like direct band gaps around 2\,eV was theoretically predicted and is assumed to take place at the K or M point in the reciprocal space \cite{qian}. Higher energy transitions cannot unambiguously be assigned. Generally there are strong resonances around 5\,eV, i.e. the maxima in the $\varepsilon_2$ spectra, which are most likely connected to even other O $2p$ to Mn $3d$ orbital transitions \cite{choi08prb77}. 
The occurrence of critical point-like direct band gaps around 2\,eV was theoretically predicted and is assumed to take place at the K or M point in the reciprocal space \cite{qian}. Higher energy transitions cannot unambiguously be assigned. Generally there are strong resonances around 5\,eV, i.e. the maxima in the $\varepsilon_2$ spectra, which are most likely connected to even other O $2p$ to Mn $3d$ orbital transitions \cite{choi08prb77}. 
 
\begin{figure*}
	\centering
		\includegraphics[width=0.95\textwidth]{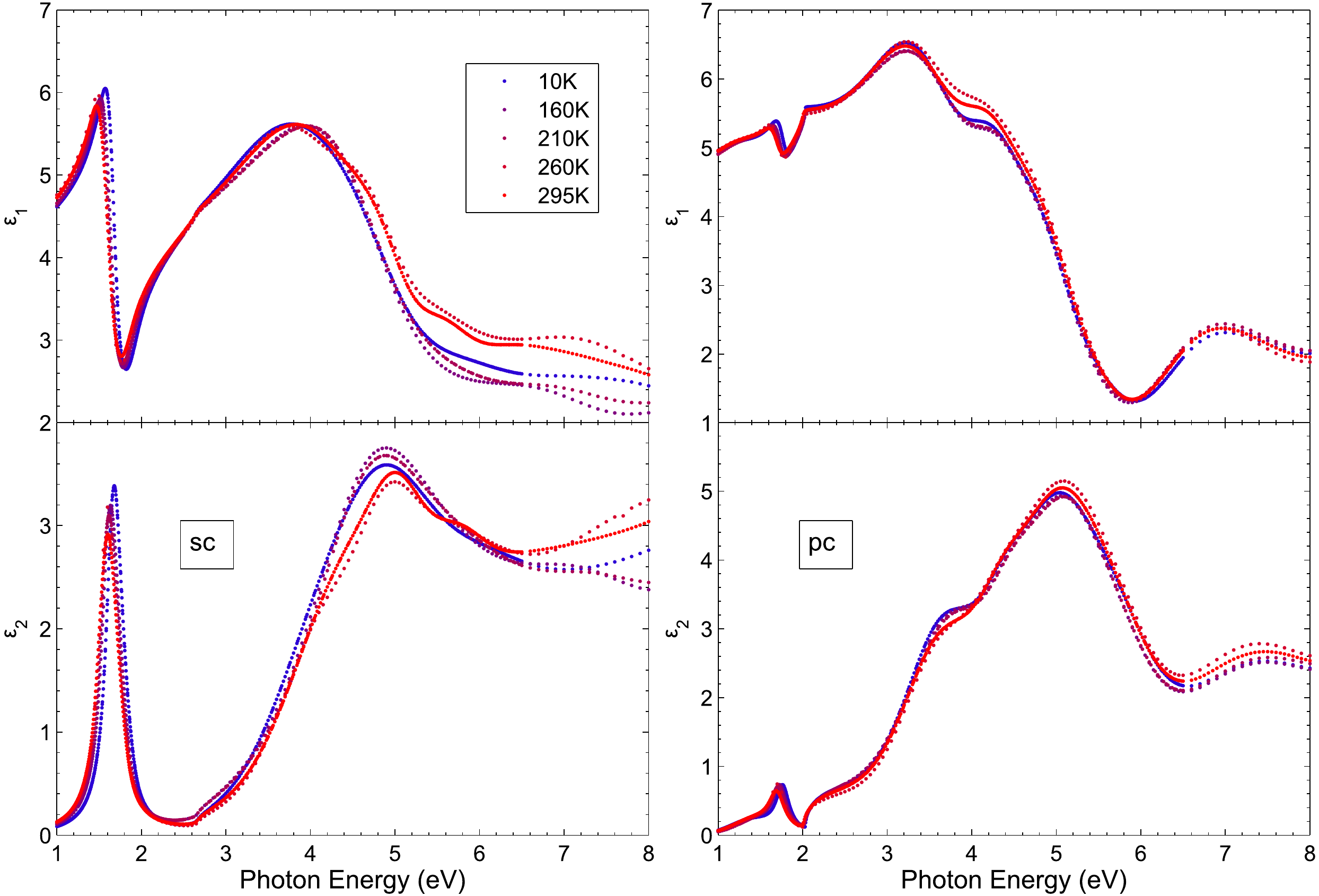}
	\caption{Real (top) and imaginary (bottom) part of the parametric model-DF shown for selected temperatures.}
	\label{fig:dfbis8eV}
\end{figure*}

The parametrization enables robust modeling at all temperatures in the spectral range from 1 to 8\,eV. As an example experimental and model-generated data for 10K is shown in fig. \ref{fig:datenanpassung}\,a). Figure \ref{fig:dfcontributions} depicts the obtained contributions to the parametric model-DF. Mean values for the found function parameters for all temperatures are listed in tab. \ref{tab:parametrization}. 
The observed dielectric functions for various temperatures are exemplarily shown in fig. \ref{fig:dfbis8eV}. 
No general temperature trend of the individual resonances could be found. 

\begin{figure*}
	\centering
		\includegraphics[width=\textwidth]{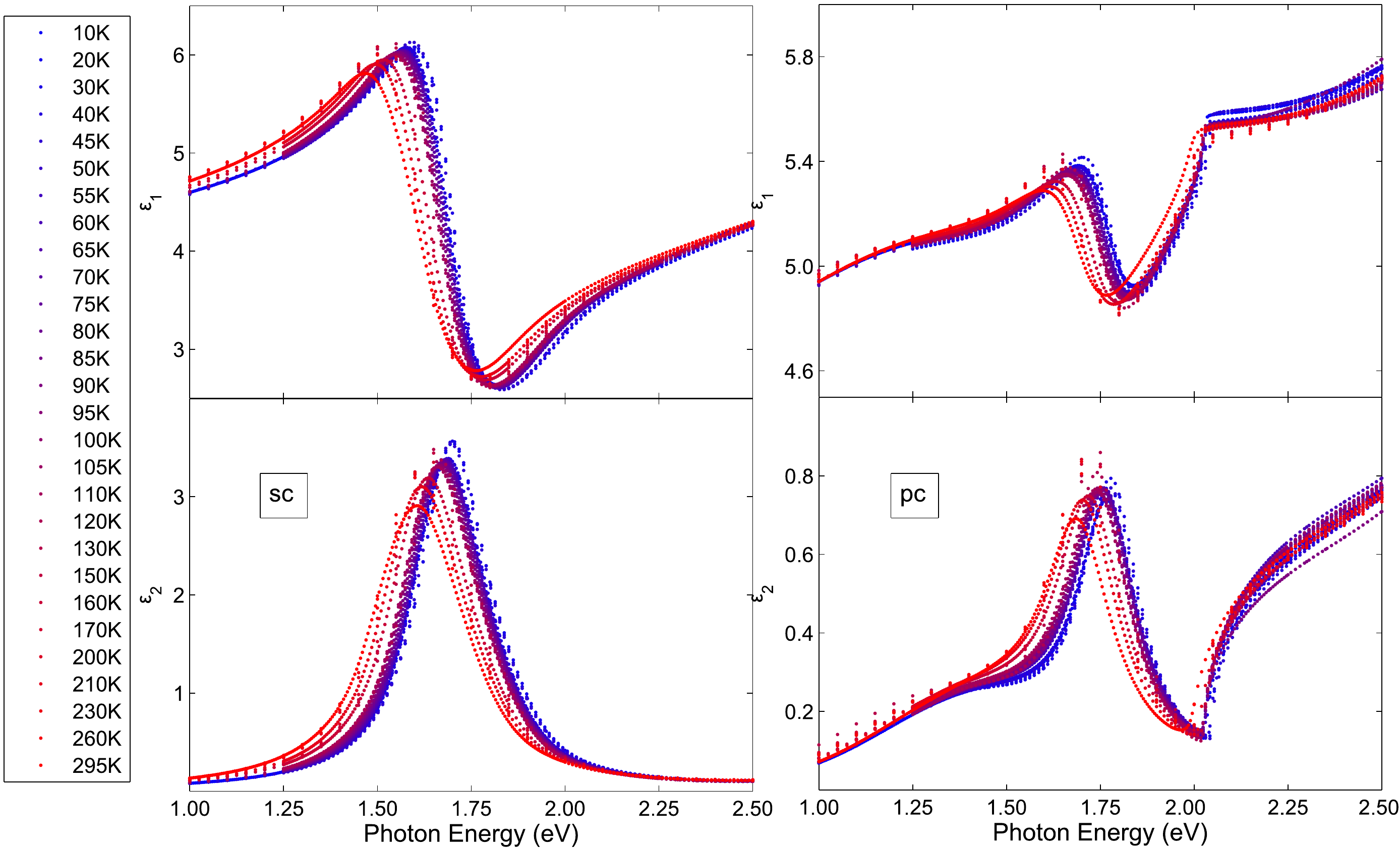}
	\caption{Real (top) and imaginary (bottom) part of the parametric model-DF in the spectral range ($1-3$)\,eV.}
	\label{fig:dfbis3eV}
\end{figure*}

\textit{Temperature dependence of the pronounced charge transfer transitions}

\begin{figure*}
	\centering
		\subfloat{
		\begin{overpic}
			[height=.375\textwidth]	{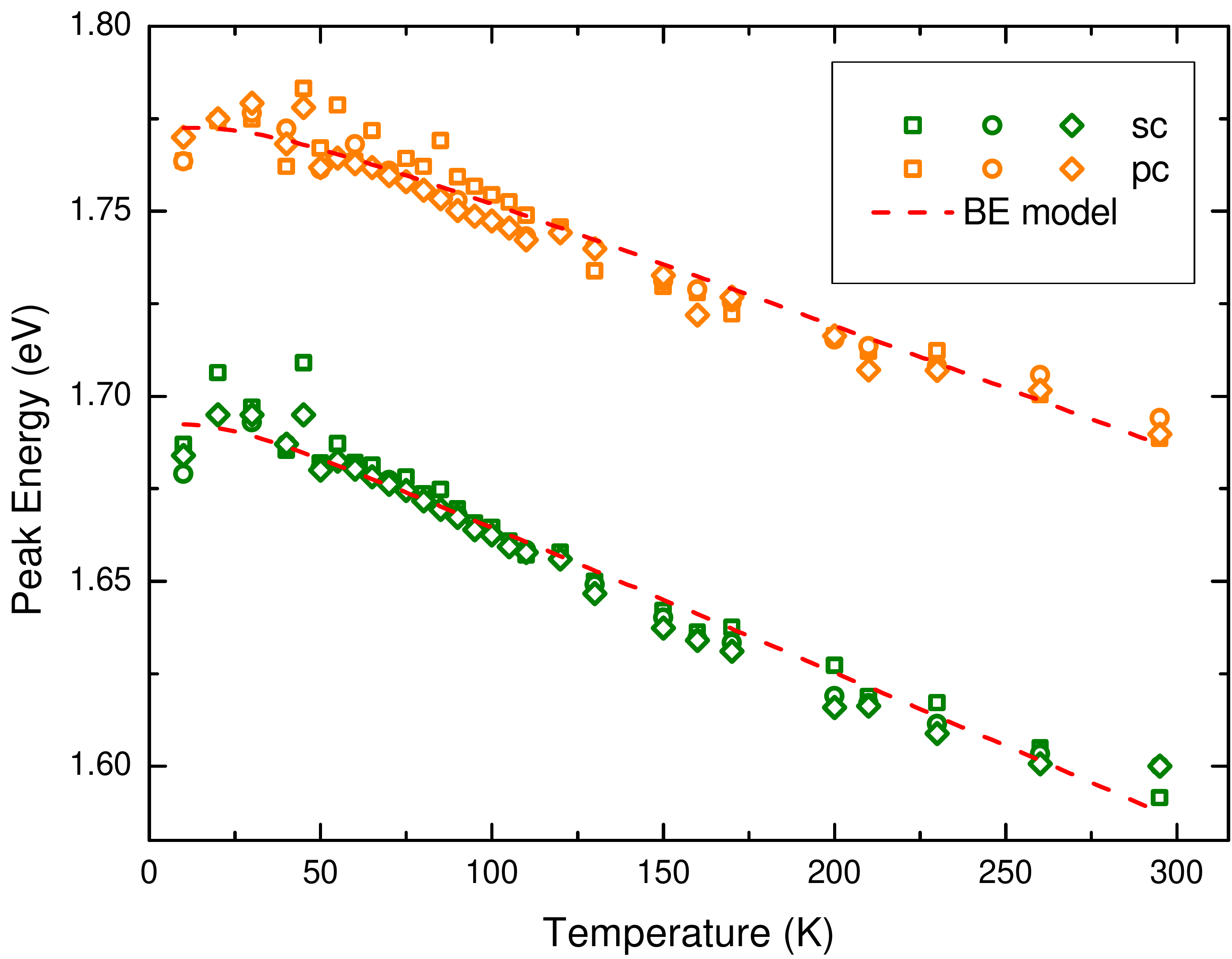}
			\put(0,0){a)}
		\end{overpic}
        }
	    \subfloat{
		\begin{overpic}
			[height=.375\textwidth]	{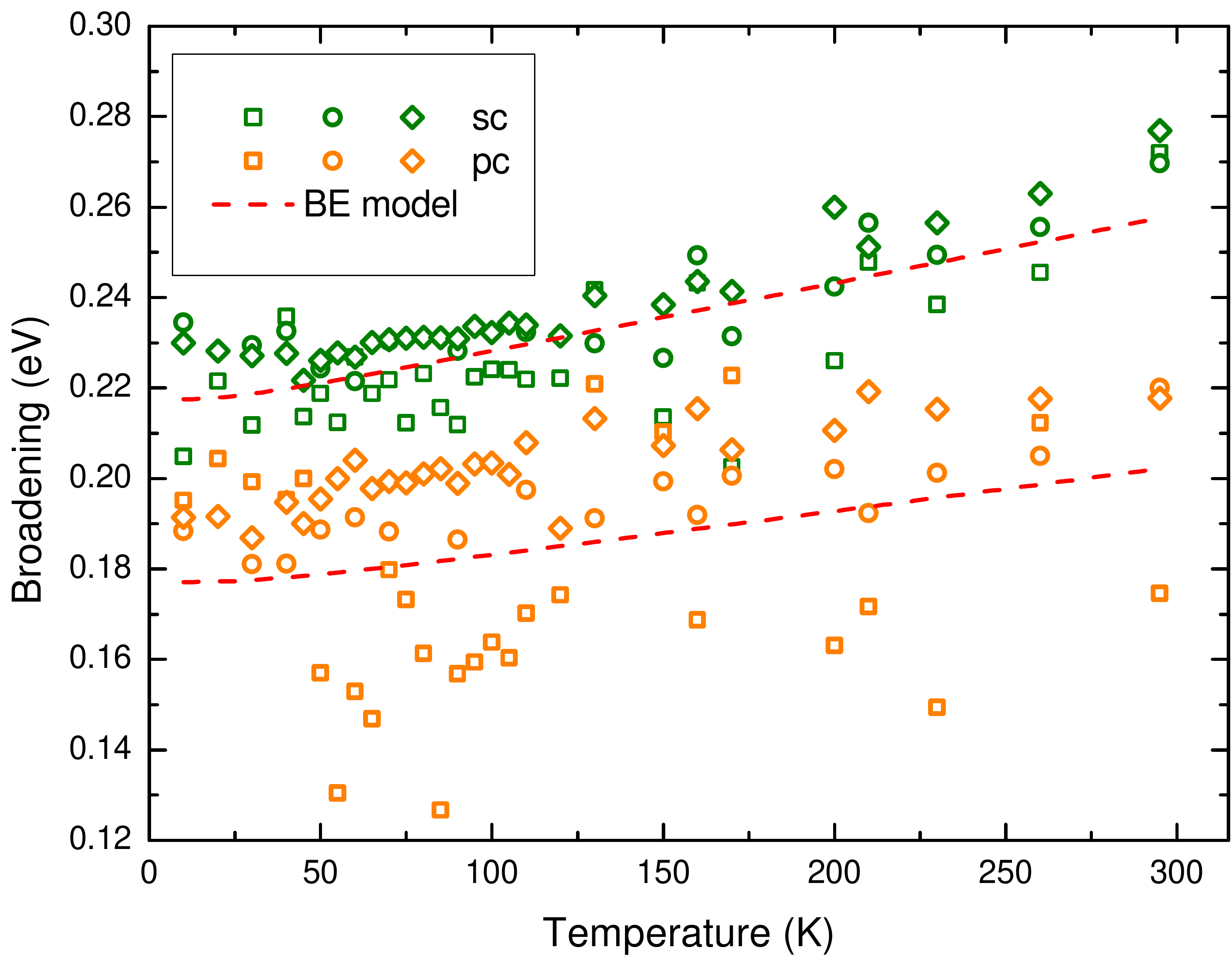}
			\put(0,0){b)}
		\end{overpic}
        } \\
	    \hspace{.1cm}
	    \subfloat{
		\begin{overpic}
			[height=.375\textwidth]	{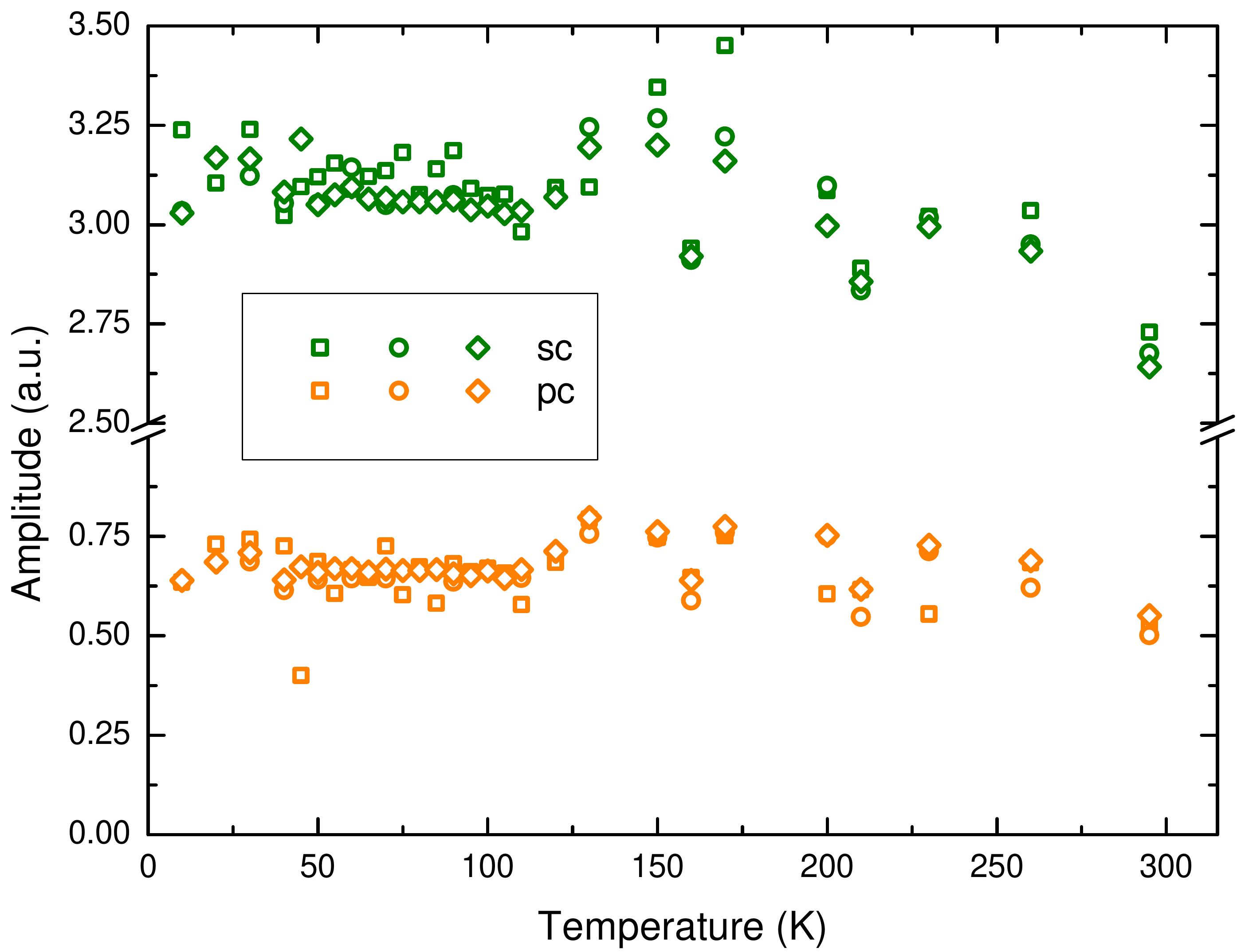}
			\put(0,0){c)}
		\end{overpic}
        }
	    \subfloat{
		\begin{overpic}
			[height=.375\textwidth]	{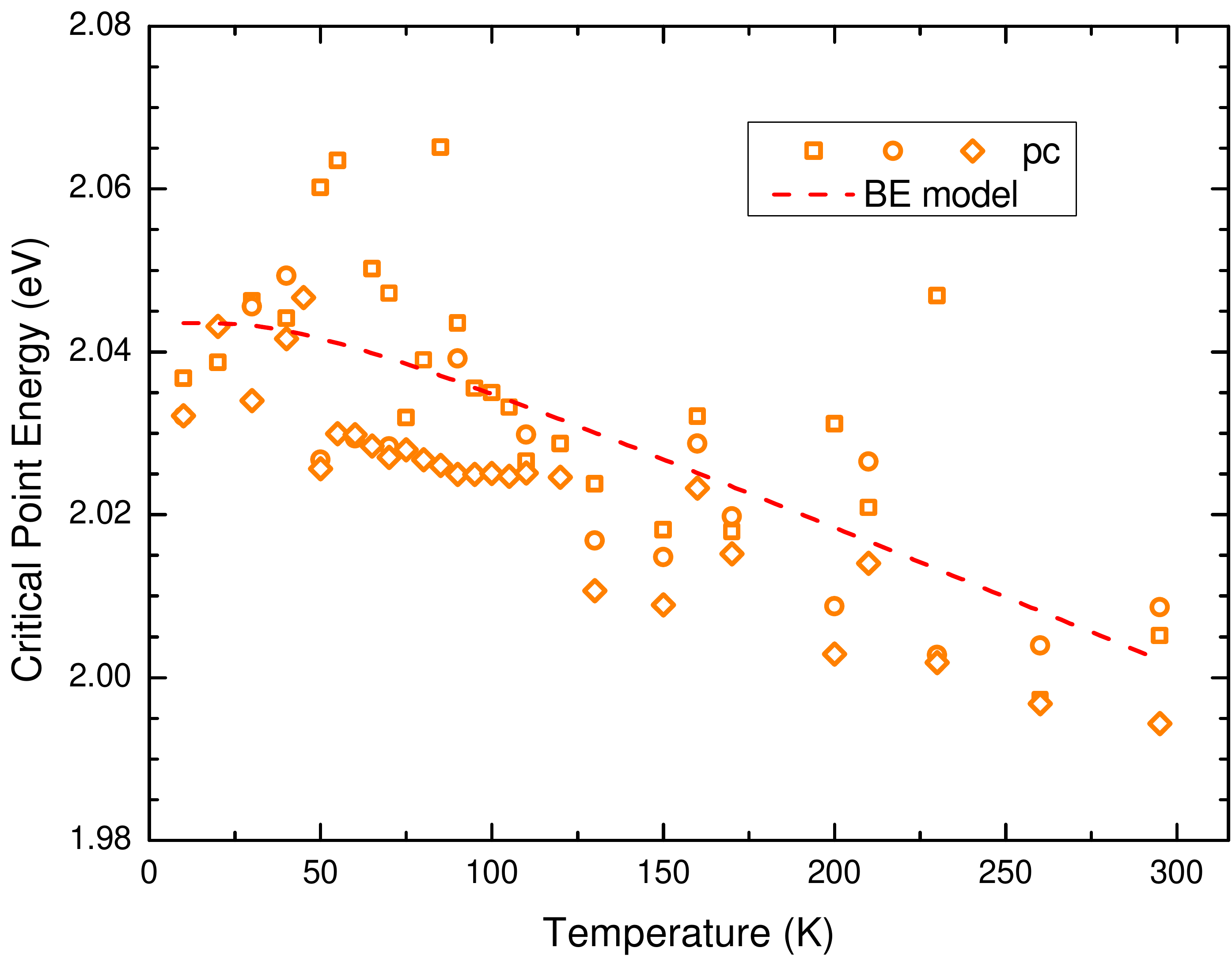}
			\put(0,0){d)}
		\end{overpic}
        }
	\caption{Temperature dependence and Bose-Einstein model approximation of the parameters of the Lorentz oscillators describing the interband charge transfer transitions: a) energies, b) broadening, c) amplitude. d): Temperature dependence of the critical point energy in pc polarization. The critical point energy in sc polarization is roughly constant at 2.6\,eV for all temperatures. Different symbols represent different evaluation series (cf. section 'Experimental details'): $\medcircle$: application of parametric model-DF to data ($1-8$)\,eV, $\meddiamond$: parametric model-DF to data ($1-3$)\,eV, $\medsquare$: numeric model-DF to data ($1-3$)\,eV and subsequent approximation by parametric model-DF.}
	\label{fig:BE}
\end{figure*}

The most pronounced features of the DF are the molecule-like peaks related to interband charge transfer transitions from the hybridized O 2$p$ to Mn $3d$ states. Figure \ref{fig:dfbis3eV} illustrates the evolution of the dielectric function with temperature in this spectral range. In order to study the temperature dependence of those main transitions in more detail, we extract peak energies $E$, broadening $\gamma$ and amplitudes as shown in fig. \ref{fig:BE}\,a)-c). The temperature dependence can be modeled using the semi-empirical Bose-Einstein model \cite{luis}: 

\begin{equation*}
\begin{aligned}
E=E_0-\frac{\alpha ~\theta}{e^{\theta/T}-1}  ~ \\
\gamma=\gamma_0+\frac{\alpha_\gamma ~\theta}{e^{\theta/T}-1} ~.
\label{eq:be-formel}
\end{aligned}
\end{equation*}
\vspace{.2cm}

Here, $E_0$, $\gamma_0$ are the values at $T=$\,0K, $\theta$ is an effective phonon temperature and $\alpha$ the effective phonon coupling constant. 
As can be seen in fig. \ref{fig:BE} the extracted values scatter a lot but the different evaluation approaches described above give consistent results. For correlation reasons we use for $\gamma (T)$ the same effective phonon temperature $\theta$ as for $E(T)$. The obtained parameters are summarized in tab. \ref{tab:be-parameter}. Remarkable are the very low effective phonon temperatures of $\theta_{sc}\approx 70$K and $\theta_{pc}\approx 90$K. This is even more striking if compared to the Debye temperature $T_D$ of YMnO$_3$ which is assumed to be rather high. Values between 350 and 500K have been reported \cite{tomuta, tachibana, lee08}. 
Consequently, this suggests also larger effective phonon temperatures in the Bose-Einstein model. For comparison, in ZnO values of $\theta\gtrsim$\,200K  and $T_D\gtrsim$\,400K were found\cite{znobuch}. An explanation for the suppressed effective phonon energy is vanishing of (at least some) phonon mode energies at the antiferromagnetic phase transition, i.e. soft mode behavior. In fact, such soft phonon modes connected to the antiferromagnetic transition are well known in perovskite-type manganites \cite{kumar,sakai,laverdiere}. 
Regarding hexagonal manganese oxides, a phonon mode connected to vibration of manganese and oxygen ions mainly within the $ab$-plane was proposed as soft mode by comparing temperature-dependent structural data \cite{zhou,lee}. 
Distinct phonon modes have directly revealed soft mode behavior at the N\'eel temperature by IR spectroscopy for h-YbMnO$_3$ \cite{liu}. 

The weak transitions accompanying the strong resonances generally show the same temperature dependence within the sensitivity of our experiment. 
For comparison we also applied the Bose-Einstein model to the critical point transition in pc polarization. See fig. \ref{fig:BE}\,d) and tab. \ref{tab:be-parameter}. The effective phonon temperature $\theta_{cp}\approx$130K is much larger than for the strong transitions, suggesting a weaker coupling of this transition to the magnetic ordering. This supports the soft mode proposal as the transitions from the hybridized oxygen to Mn-$3d$ states can be expected to reveal one of the closest connection to the magnetic correlation compared to any other transition. For sc polarization the found critical point energy scatters too much to analyze it further. Changes of the oscillator amplitudes with temperature do not reveal a clear tendency either (cf. fig. \ref{fig:BE} c)). 

Regarding anisotropy we observe the soft mode coupling in both directions, sc and pc, but it is significantly larger (i.e. stronger suppression of the effective phonon temperatures in the Bose-Einstein model) for the polarization perpendicular to the $c$-axis. 
This is consistent with the interpretation of structural and thermal conductivity data claiming a displacement of the Mn ion in the $ab$-plane as origin of phonon softening \cite{zhou}. 
However, while this thermal conductivity of h-YMnO$_3$ was also affected in both directions, the antiferromagnetic phase transition was only observed perpendicular but not parallel to the $c$-axis in the (low frequency) dielectric constant \cite{katsufuji}. 
Understanding the polarization dependence is not trivial as it requires knowledge of each individual phonon mode and its coupling to the respective electronic transition. E.g. for YbMnO$_3$ the soft mode behavior was only observed for A$_1$ phonon modes (pc), which influence inter-plane Mn-Mn coupling, while no soft mode character was found for E$_2$ modes (sc), which however revealed direct magnetic coupling \cite{liu}. 
In conclusion, we find impact of the antiferromagnetic phase transition on the electronic transitions in both polarizations, sc and pc. Its coupling via soft phonon modes varies between the different electronic resonances. 

\begin{table*}
	\centering
		\caption{Bose-Einstein model parameters.}
		\vspace{.5cm}
		\begin{tabular}{c c c c c c c}
			 & $E_0$ & $\alpha$ & $\theta$ & $E_\theta$ & $\gamma_0$  & $\alpha_\gamma$\\
			polarization & eV & meV/K & K & meV & meV  & meV/K \\
			\hline
			sc & $1.692(3)$ &	$0.399(16)$	& $67.5(26.4)$ & $5.85(2.29)$ &	$217.5(2.1)$	& $0.153(17) $ \\
			pc & $1.773(2)$ &	$0.341(17)$ & $93.2(29.3)$ & $8.08(2.54)$ &	$177.1(3.9)$ & $0.100(17)$ \\
		\end{tabular} 
		\vspace{.25cm}
		\\
		\begin{tabular}{c c c c c}
		    & $E_0^{cp}$ & $\alpha^{cp}$ & $\theta^{cp}$ & $E_\theta^{cp}$\\
			polarization & eV & meV/K & K & meV\\
			\hline
			pc & $2.044(4) $ & $0.175(43)$ &	$126(133)$ & $11.0(11.5)$ \\
		\end{tabular}
	\label{tab:be-parameter}
\end{table*}

\begin{figure}
	\begin{center}
		\includegraphics[width=0.5\textwidth]{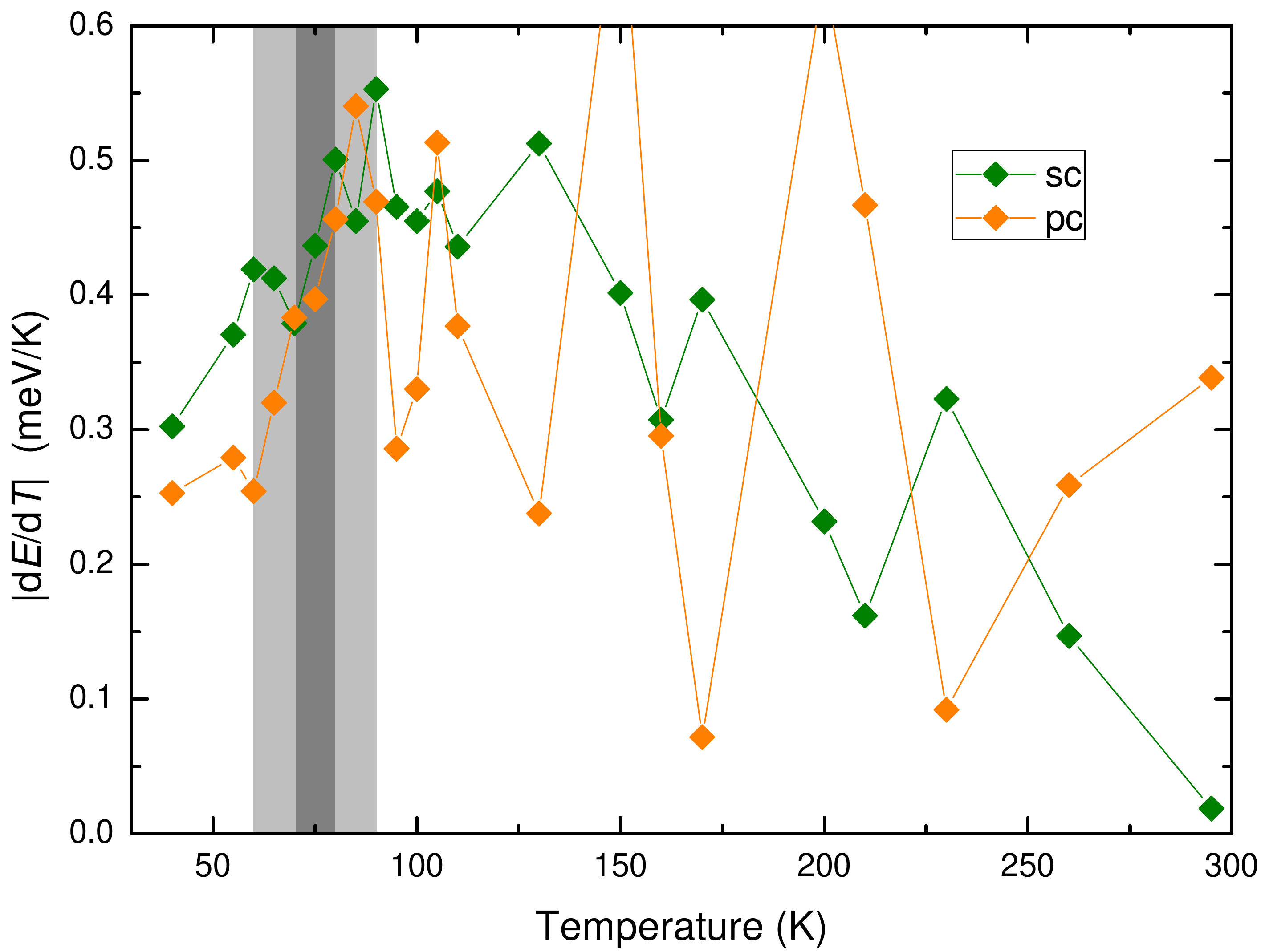}
    \end{center}
	\caption{Modulus of the derivative of the transition energies of the strong charge transfer transitions. The gray area marks the range of literature values for the N\'eel temperature. The data was taken from the '$\meddiamond$' series (see caption of figure \ref{fig:BE}).}
	\label{fig:deriv}
\end{figure}

As an alternative to the Bose-Einstein model approximation we investigated the first derivative $\frac{\mathrm{d}E}{\mathrm{d}T}$ of the transition energies with respect to the temperature for the main charge transfer transitions. It is shown in fig. \ref{fig:deriv}. Indeed, the slope as a function of temperature reveals a maximum near the N\'eel temperature, similar to what was found for other h-$R$MnO$_3$ materials in sc polarization \cite{choi08prb78}. We also observe this anomaly in both polarizations, parallel and perpendicular to the crystallographic $c$-axis. 
Such effect is not captured by the Bose-Einstein model and again strongly indicates a magneto-electric coupling mediated through a strong charge-spin interaction. 

\textit{Photoluminescence}

\begin{figure}
	\begin{center}
	 \quad \includegraphics[width=.475\textwidth]{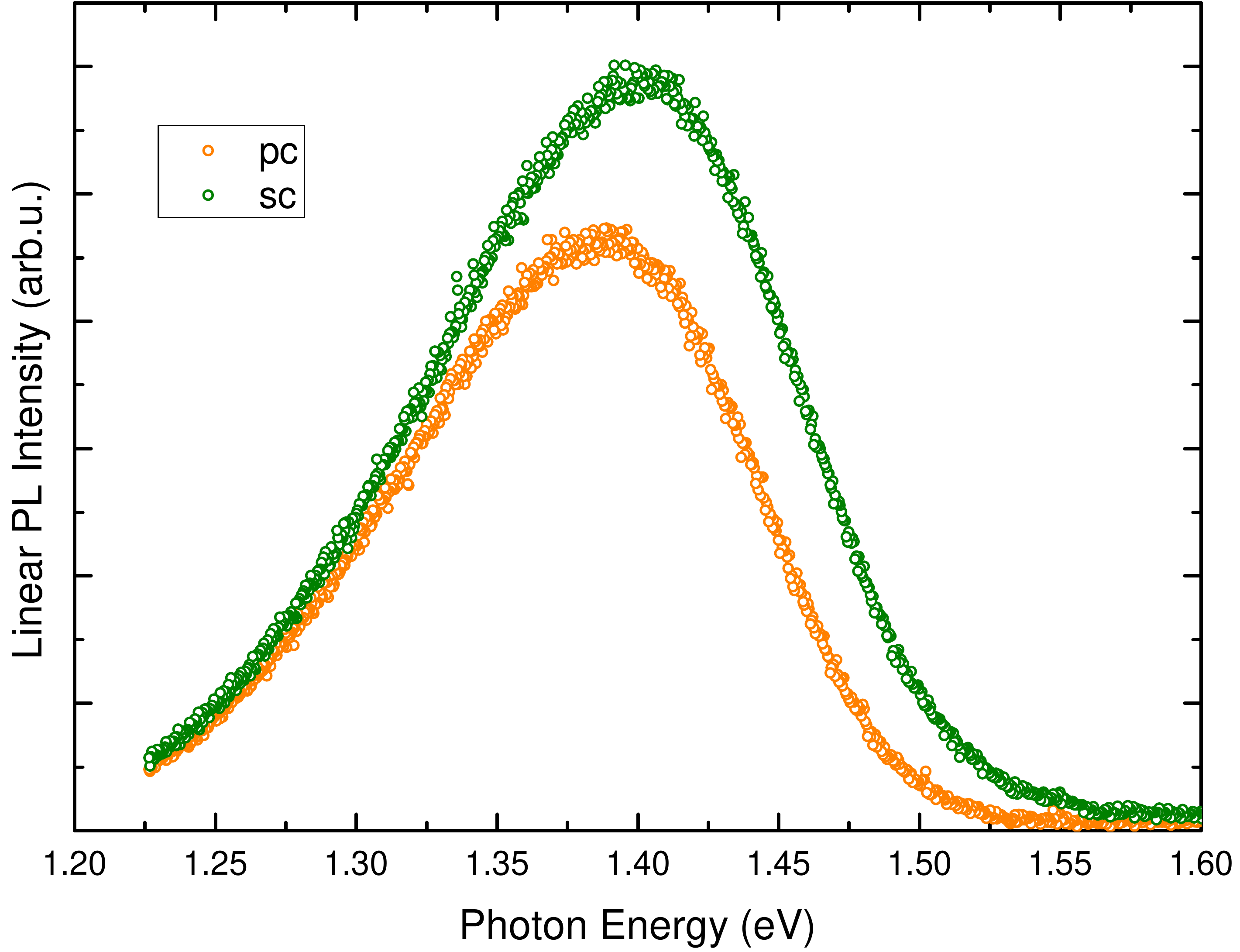}
	\end{center}
	\caption{Photoluminescence intensity at 10K with 633nm excitation (HeNe laser).}
	\label{fig:pl}
\end{figure}

Low temperature photoluminescence (PL) measurements yield broad emission peaks around 1.4\,eV (fig. \ref{fig:pl}), similar to earlier reported PL spectra from h-YMnO$_3$ single crystals \cite{takahashi}. 
As origin of the emission, the strong charge transfer transition involving excitation of a phonon and two magnons was proposed. This also explains the large broadening of the signal due to the wide two-magnon continuum. 
However, in comparison with the uniaxial DF it is not clear if the origin is the main transition or a (in the DF) weaker one. Especially the energetic order of the emission peaks with respect to their polarization is not the same as in the DF. While in the DF the pc polarization reveals the higher energy, in the PL spectrum the sc polarization does. 
It should further be mentioned that the sc polarized PL spectrum of a thin h-YMnO$_3$ film looks completely different: Sharp features at 1.7\,eV and broad emission around 1.55\,eV were observed \cite{nakayama}. Furthermore the emission increased abruptly below $T_N$, a behavior we could not observe. Most likely interface effects between the substrate and the film play an important role in the ferroelectric-antiferromagnetic thin films.

\section*{Conclusion}

An oscillator function-based dielectric function for h-YMnO$_3$ was developed and temperature-dependent generalized spectroscopic ellipsometry was used to deduce the anisotropic dielectric function of YMnO$_3$ in the spectral range ($1-8$)\,eV for temperatures from 10K to room temperature. 
The temperature dependence of energy and broadening of the very pronounced Mn ion-related charge transfer transition peaks can be described by a Bose-Einstein model only if very low effective phonon energies (8\,meV and smaller) are assumed. This gives a hint to soft mode behavior connected to the antiferromagnetic phase transition at the N\'eel temperature. The phase transition can also be observed in the derivative of the resonance energies with respect to the temperature. Hence, coupling between magnetic order and the electronic structure can be revealed by optical spectroscopy in the visible spectral range. 
For energetically higher electronic resonances no general temperature trend could be found. 
Furthermore, the spectral weight does not considerably change with temperature neither energetically nor between the polarizations. 
Finally we compare the low temperature dielectric function to photoluminescence spectra and find that the origins of the emission are not necessarily the main charge transfer transitions.

\section*{Acknowledgements}
We acknowledge Tom Michalsky for carrying out the photoluminescence measurements. This work was funded by the Sonderforschungsbereich (SFB) 762 'Funktionalit\"at oxidischer Grenzfl\"achen' (INST271/240-2, A8 and B3). SR acknowledges support by the graduate school BuildMoNa. We also thank Helena Franke and Chris Sturm for help with the manuscript.

\bibliography{ymo_tempabh_df}{}
\bibliographystyle{unsrt}

\end{document}